# Quantum and optics effects in dense plasmas with medium temperatures with implication to stellar plasmas

**Y. Ben-Aryeh**,

Technion-Israel Institute of Technology, Physic Department, Israel, Haifa, 32000

Email:   phr65yb@technion.physics.ac.il

The optical properties of plasmas with high densities and medium temperatures are analyzed by the use of a free electron model with Fermi-Dirac statistics. For the present collisional plasma the frequency of electron-ion collision is very large relative to the optical and infra-red frequencies. A quantum mechanical equation for the frequency of collisions is developed by the use of Fermi-Dirac statistics and Rutherford scattering theory. The validity of the Rutherford scattering theory is discussed. The influence of many weak collisions is taken into account by a Coulomb logarithmic function. The present analysis might have implication to stellar plasmas with medium temperatures for which Fermi-Dirac statistics is used. The relations between the present analysis and the stabilities of stars plasmas are discussed. The ratio between the radius and mass of star plasmas with the present densities and that of a typical white dwarf are discussed.

## 1. Introduction

The aim of the present work is to treat quantum and optics effects in plasmas with high densities and medium temperatures which are related to Fermi-Dirac statistics (Aschcroft & Mermin 1976). While Fermi-Dirac statistics was applied to white dwarfs (Kippenhahn & Weigert 1990) which have very high densities, in the present work, we study plasmas with densities and temperatures which are lower than those in white dwarfs. Optical properties of plasmas are treated in various books (Ginzburg 1961; Chen 1990; Piel 2010 & Fitzpatrick 2015). Usually one uses Boltzmann statistics for describing the statistics of ionized electrons, but we treat here the plasmas under the conditions by which quantum effects become important and Boltzmann statistics is exchanged to Fermi-Dirac statistics. Fermi-Dirac statistics was used also in the field of semi-conductors (Blakemore 1962 & Seeger 1991). By taking into account that the ratio $M_i/m$ is very large, where $M_i$ and $m$ are the ion and the electron mass, respectively, the plasma properties are described mainly by the electrons which are very mobile relative to the ions. The electron de-Broglie wave length is given by $\lambda_{DB} = h/mv$. Here $h$ is the Planck constant, $m$ and $v$ are the mass and velocity of the electron, respectively, and the quantum effects become important when the de-Broglie wave length is large relative to the average distance between the ionized electrons. So that if $\lambda_{DB} \gg n^{-1/3}$ (where $n(m^{-3})$ is the number of electrons per unit volume, and $n^{-1/3}(m)$ is the average distance between the electrons) the Boltzmann statistics breaks down and it should be exchanged to Fermi-Dirac statistics.

In the present work we compare the preset analysis with the works of other authors. Usually one would think that recombination (both of atoms and molecules) would be a dominant effect in the present region. Elementary calculations using the Saha model are then made for the dissociation and ionization of the atoms and molecules. Such calculations are usually based on Boltzmann statistics. The approach for stellar atmospheres is, however, different. In the analysis of the stability of stellar plasmas it is assumed that the pressure $P$ of the plasma and the electrons velocities $v$ are dependent on the density of electrons and ions and their direct dependence on temperature $T$ is neglected (Chandrasekhar 2010 & Eddington 1959). The justification for such an approximation follows from the fact that the pressure of the dense fermionic plasma is a purely quantum mechanical effect related to the uncertainty and Pauli exclusion principles. We compare the present analysis with those of white dwarfs which are made for much higher densities and temperatures. The optical properties of the present plasmas are investigated,

The electron dielectric permittivity has been calculated in various works (Brantov, Bychenkov, Rozmus & Capjack 2004). Dielectric function in the collisional degenerate plasma was derived by using a Fokker-Planck model (Nersisyan, Veysman, Andreev & Matevosyan 2014). Scattering of electrons around atoms in degenerate matter was studied (Kuchiev & Johnson 2008). Such problems were studied in many works but the results become quite complicated. We would like to develop here a simple relaxation-time approximation based on previous works (Mermin 1970), in which the collisions relax the electrons to their Fermi-Dirac distribution and the frequency $\omega$ is replaced by $\omega - iq$ where $q$ is the frequency of collisions. Free electron model was developed (Maier 2007) in which the effect of the collisions on the plasma optical properties is taken into account but we would like to generalize such model for the degenerate plasma. Calculations of dense-plasma opacities were published recently (Piron & Blenski 2018), but as we treat plasma with different conditions the preset analysis is different.

For the high densities plasmas and medium temperatures the electron-ion collision frequency might be very large relative to the optical or infra-red frequencies. In order to estimate the electron-ion collision frequency we start from the classical Rutherford scattering theory. This classical expression averaged over energies with the Maxwell distribution leads to the well-known Spitzer formula for the conductivity (Meyer-ter-Vehn, Tronnier & Cang Y. 2008) but such analysis does not apply for the present analysis for the degenerate plasma. Analytical expression for inverse stimulated bremsstrahlung has been derived for slow electrons under Coulomb scattering (Krainov 2000). The use of Rutherford scattering theory is related to Born approximation which is valid under the condition $\frac{Ze^2}{4\pi\varepsilon_0 \hbar v} < 1$, where $Z$ is the atomic number, $e$ the electronic charge, $\hbar$ the Planck constant (divided by $2\pi$) and $v$ the electron velocity (Marcuse 1962). (In the present work we use the MKS (*Met.Kg.Sec.*) unit system). One should take into account that the velocity of electrons in the Fermi-Dirac statistics is a function of the density of electrons and not their temperature. Taking this into account we find that the use of Rutherford scattering theory is a fairly good approximation for the density of electrons $n = 10^{31}$ ($m^{-3}$) (treated as an example in the present work) and for higher densities but such method breaks down for lower densities. The classical Rutherford collision frequency is proportional to $1/v^3$ where $v$ is the electron velocity. We obtain new quantum mechanical equation for the collision frequency, in the region where Rutherford scattering theory is valid, by averaging this function over the Fermi-Dirac distribution. Since the Coulomb interactions are of long range many collisions with small changes in energy contribute to the collision frequency by amount which is larger than that of the strong collisions. Following this idea we develop a certain estimate for the Coulomb logarithmic function.

The present paper is arranged as follows: In section 2 we describe the quantum effects occurring in plasmas by relating the density of electrons to the Fermi-Dirac distribution in the non-relativistic region. Although such result is well known we would like to present it for the consistency in later calculations. In section 3, I develop quantum mechanical equations for the collisions frequencies which take into account strong collisions. The collision frequency is obtained by integration over Fermi-Dirac distribution. Such integral leads to Coulomb factor $\ln(p_F / p_{\min})$ where its upper limit is given by the Fermi-electron-momentum and its lower limit $p_{\min}$ is related to the effect of many collisions with small changes in energy. The equations for the optical properties of cold plasmas with medium densities are developed by using free electron model under the condition that the quantum mechanical collision frequency is much larger than the optical or infra-red frequencies. In section 4 we demonstrate our analysis by numerical calculations, for a certain example. In section 5 we find relations between our analysis and stars plasmas stabilities. We estimate the relation between the radius and mass of a star with the preset conditions with those of a typical white dwarf. In section 6 we summarize our results and conclusions.

## 2. The density of electrons at medium temperatures according to Fermi-Dirac statistics with non-relativistic approximations

The total number of electrons, $N_{FD}$ in the non-relativistic approximation is given by the Fermi-Dirac equation (Aschroft and Mermin 1976):

$$N_{FD} = 2\frac{4\pi V}{h^3}\int_0^\infty f(x,p)p^2 dp \quad ; \quad f(\vec{x},\vec{p}) = \frac{1}{\exp\left[(\frac{p^2}{2m}-\mu)/kT+1\right]} \quad . \tag{1}$$

Here $V$ is the volume of our system and $(N/V)_{FD} = (n)_{FD}$ is the number of electrons per unit volume. $f(\vec{x},\vec{p})$, is the Fermi-Dirac probability density function, i.e. the probability of finding a particle at position between $\vec{x}$, and $\vec{x}+d\vec{x}$, and its 3-momentum between $\vec{p}$, and $\vec{p}+d\vec{p}$, $E = p^2/2mkT$ is the electron non-relativistic energy in units of $kT$ where $k$ is the Boltzmann constant, $T$ is the temperature, $m$ is the electron mass and $\mu$ is the chemical potential, $h^3$, is the volume of a quantum cell in phase space where $h$ is the Planck constant and the factor 2 follows from the two spin states of the electron. The chemical potential $\mu$ is to be chosen in the particular problem in such a way that the total number of electrons comes out correctly-that is equal-$N$. For degenerate fermions with $\mu/kT = \eta > 1.5$ the function $f(\vec{x},\vec{p})$ is approximated as

$$f(\vec{x},\vec{p}) = f(E) = \begin{cases} 1 & \text{for } E = \frac{p^2}{2m} \leq E_F \\ 0 & \text{for } E = \frac{p^2}{2m} > E_F \end{cases} \quad ; \quad E_F \equiv \frac{p_F^2}{2m} \equiv \mu \quad ; \quad \frac{\mu}{kT} \equiv \eta \gg 1.5 \quad . \tag{2}$$

By substituting the approximation (2) into Eq. (1) we get:

$$n_{FD} = (N/V)_{FD} = \int_0^{p_F} \frac{8\pi p^2}{h^3} dp = \frac{8\pi p_F^3}{3h^3} \quad , \quad p_F = h\left(\frac{3n}{8\pi}\right)^{1/3} = \sqrt{2mE_F} \tag{3}$$

The approximation (2) for the Fermi wave vector $p_F$ is referred often as the zero-temperature Fermi-Dirac distribution but such definition might be misleading as the condition in (2) might be valid for relatively high temperatures, depending on the magnitude of the chemical potential (Seeger 1991). The properties of the plasmas depend on the density of the electrons and not on their temperature for cases for which (2) is satisfied.

## 3. The optical properties of cold plasma with medium densities in the non-relativistic region

By using free electron model the dielectric constant for monochromatic electromagnetic waves propagating in plasma is given by (Maier 2007):

$$\varepsilon = \left(1 - \frac{\omega_p^2}{\omega^2 + iq\omega}\right)\vec{E} \quad ; \quad \omega_p^2 = \frac{ne^2}{\varepsilon_0 m} \quad . \tag{4}$$

Here, $\varepsilon$ is the complex dielectric constant, $\omega$ is the light frequency, $q$ is the collision frequency, $e$ the electron charge and $\omega_p^2$ is the plasma frequency of the free electron gas. The real and imaginary of the complex dielectric constant: $\varepsilon = \varepsilon_1 + i\varepsilon_2$ are given by

$$\varepsilon_1(\omega) = \left(1 - \frac{\omega_p^2}{\omega^2 + q^2}\right) \quad ; \quad \varepsilon_2(\omega) = \frac{\omega_p^2 q}{\omega(q^2 + \omega^2)} \quad . \tag{5}$$

We treat the plasma, under the condition $q \gg \omega$ by which we get:

$$\varepsilon_1(\omega) = \left(1 - \frac{\omega_p^2}{q^2}\right) \quad ; \quad \varepsilon_2(\omega) = \frac{\omega_p^2}{\omega q} \quad . \tag{6}$$

We can use the relation

$$\sqrt{\varepsilon} = \sqrt{\varepsilon_1 + i\varepsilon_2} = n + i\kappa \quad . \tag{7}$$

Here $n$ is the index of refraction and $\kappa$ is the extinction coefficient. The index of refraction $n$ and the extinction coefficient $\kappa$ are related to $\varepsilon_1$ and $\varepsilon_2$ by (Maier 2007):

$$n^2 = \frac{\varepsilon_1}{2} + \frac{1}{2}\sqrt{\varepsilon_1^2 + \varepsilon_2^2} \quad ; \quad \kappa = \frac{\varepsilon_2}{2n} \quad . \tag{8}$$

The extinction coefficient $\kappa$ is linked to the absorption coefficient of Beer's law (describing the exponential attenuation of the intensity of a beam propagating in a distance $d$ through the medium via the relation $I(d) = I_0 e^{-\alpha d}$) by

$$\alpha(\omega) = \frac{2\kappa\omega}{c} = \varepsilon_2 \frac{\omega}{cn} \simeq \frac{\omega_p^2}{cqn} \tag{9}$$

We find according to (6) that the real part of the dielectric constant $\varepsilon$ tends to the value 1 under the conditions: $q^2 \gg \omega^2$ ; $q^2 \gg \omega_p^2$ and the imaginary value of $\varepsilon$ tends zero under the conditions: $q^2 \gg \omega^2$ ; $\omega q \gg \omega_p^2$. Under such conditions the index of refraction tends to the value 1 (or in the order of 1) and then by using (9) $\alpha(\omega)$ becomes very small under the condition

$$\omega_p^2 \ll qc \quad . \tag{10}$$

This equation is satisfied under the condition that $q$ is extremely large and the density of electrons is smaller than those assumed in white dwarfs but due to medium temperatures the Fermi-Dirac statistics should be used..

The critical parameter in the present optical analysis is the estimation of the parameter $q$ representing the frequency of collision. As we are interested in estimating order of magnitudes for the collision frequency $q$ we simplify the calculation by assuming plasma with one ionic component, with atomic number $Z_i$. By taking in account that when the electron is colliding with ion it is deflected by long range Coulomb potential, we estimate the value of $q$ by electron-ion collisions. The collision for frequency for electron ion collisions is obtained by following Rutherford scattering theory (Ginzburg 1961; Chen 1990; Pitzpztrik 2015) and it is given by:

$$q = \frac{nZ_i^2 e^4}{4\pi\varepsilon_0^2 m^2 v^3} = \frac{nZ_i^2 e^4 m}{4\pi\varepsilon_0^2 p^3} = Z_i^2 n \cdot 0.6091 \cdot 10^{-84} \cdot \frac{1}{p^3} \quad (\text{sec}^{-1}) \quad . \tag{11}$$

There is a divergence in this equation for $p \to 0$ but this divergence is eliminated later by taking into account many weak collisions effects. We calculate the averaged value of the frequency of collision, by integrating it over the Fermi-Dirac distribution which takes into account quantum effects, given by:

$$q_{FD} = C\frac{2}{h^3}\int_0^\infty \frac{1}{p^3} f(x,p) d^3p \quad ; \quad C = \frac{Z_i^2 e^4 m}{4\pi\varepsilon_0^2} = Z_i^2 \cdot 0.6091 \cdot 10^{-84} \quad (\text{sec}^{-1}) \quad . \tag{12}$$

One should take into account that the expression $\frac{2}{h^3}\int_0^\infty f(x,p) d^3p$ represents the Fermi-Dirac distribution for the number of electrons. This distribution should be multiplied by $\frac{C}{p^3}$ where the Fermi-Dirac averaging is done only on $1/p^3$. Using the approximation (2) and inserting the numerical values for $C$ and $h^3$ we get:

$$q_{FD} = C\frac{2}{h^3}\int_{p_{\min}}^{p_F} \frac{4\pi p^2}{p^3} dp \simeq Z_i^2 5.263 \cdot 10^{16} \int_{p_{\min}}^{p_F} \frac{1}{p} dp \simeq Z_i^2 \cdot 5.264 \cdot 10^{16} \ln\left(\frac{p_F}{p_{\min}}\right) \quad (\text{sec}^{-1}) \quad . \tag{13}$$

Here we inserted in the integration of (13) the lower bound $p_{\min}$ which is related to the effect of many weak collisions and which prevents the divergence of this equation. The Coulomb logarithmic function $\ln\left(\frac{p_F}{p_{\min}}\right) \equiv \ln(\Lambda)$ is usually calculated by Boltzmann statistics for various kinds of plasma (Chen 1990), but here we need to use this function for Fermi-Dirac statistics.

In the present work we estimate the Coulomb logarithmic function by using the following interpretation. The momenta $p_F$ and $p_{\min}$ in (13) may represent changes in electron momentum for strong and weak collisions, respectively. The classical equation (11) for the frequency of collisions $q$ is obtained by assuming strong elastic collisions where the change in the electron momentum is given by $\Delta(mv)$ which is equal approximately to the electron momentum $mv$ (Chen 1990). Such calculation does not take into account the effect of many weak collisions which involve small changes of energy. The logarithmic function $\ln\left(\frac{p_F}{p_{\min}}\right)$ with parameter $\left(\frac{p_F}{p_{\min}}\right)$, can be exchanged into a logarithmic function which includes as parameter the ratio between the Fermi energy and a certain minimal energy change. This can be done by following the relation: $\ln\left(\frac{p_F}{p_{\min}}\right) = \frac{1}{2}\ln\left(\frac{p_F}{p_{\min}}\right)^2 = \frac{1}{2}\ln\left(\frac{p_F^2/2m}{p_{\min}^2/2m}\right)$. This effect is taken into account in $q_{FD}$ of (13) by the increase of $q_{FD}$ inversely proportional to $p$ with a low bound $p_{\min}$. The change of electron energy in the collision is bounded quantum mechanically by the energy $\hbar \omega$ of one photon, where $\omega$ is the frequency of the monochromatic electromagnetic field. Following these considerations we can estimate the Coulomb logarithmic function as

$$\ln(\Lambda) \simeq \ln\left(\frac{p_F^2/2m}{\hbar \omega}\right) \quad . \tag{14}$$

It is interesting to note that a similar logarithmic Coulomb function was obtained in calculations for stimulated emission of bremsstrahlung (Marcuse 1962).

Here the contribution of many weak collisions is given, approximately by the logarithm of the ratio between the Fermi energy and the energy of one photon. Since this ratio enters only as a parameter in a logarithmic function the present results do not depend critically on this parameter.

By calculating the velocity $v$ according to (3) and by substituting the collision frequency $q$ of (13) with the Coulomb logarithmic function (14) into (6-10) we get the optical properties of the plasma. One should, however, take into account that the present use of the Rutherford-Born scattering theory is valid only for densities in the order of $n = 10^{31}$ or higher, satisfying the Rutherford validity condition (Marcuse 1962 : Krainov 2000):

$$\frac{Ze^2}{4\pi\varepsilon_0 \hbar v} < 1 \quad . \tag{15}$$

In the next section we demonstrate such calculations for $n = 10^{31} \left(m^{-3}\right)$. Also in later section we relate such analysis to the stability and properties of star plasmas with high densities and medium temperatures, and compare the results with those of white dwarfs which have higher densities and temperatures.

**4. Numerical calculations**

We demonstrate the optical properties of plasma with medium temperatures by making the following calculations in an example for which: $n = 10^{31} \, (m^{-3})$. Using (3), we get for this example

$$p_F \simeq 7.03 \cdot 10^{-24} \, \left(kg.met.\sec^{-1}\right) \quad . \tag{16}$$

The Fermi-Dirac statistics is valid under the condition given in (2) by

$$\frac{p_F^2}{2mkT} \simeq \frac{1.96 \cdot 10^6}{T} \gg 1.5 \quad ; \quad T \ll 1.31 \cdot 10^{6, 0}K \quad . \tag{17}$$

For checking the condition for the Born-Rutherford approximation (15) we use the equality:

$$\frac{Ze^2}{4\pi\varepsilon_0 \hbar v} \simeq 0.28 \cdot Z_i \quad . \tag{18}$$

Then we find that the condition (15) is satisfied for hydrogen but for oxygen or carbon it can be used only as a rough approximation. It can however be improved for densities which are larger than: $10^{31} \, (met.^{-3})$. Assuming in this example an optical frequency $\omega = 10^{14}$ then the Coulomb logarithmic function of (14) is given by

$$\ln\left(\frac{p_F^2/2m}{\hbar\omega}\right) \simeq \ln(2.57\cdot 10^3) \simeq 7.85 \quad ; \quad \left(\omega = 10^{14}\right) \ . \tag{19}$$

The frequency of collisions in this example is given by using (13) and the relation (19) as:

$$q_{FD} \simeq Z_i^2 \cdot 5.264\cdot 10^{16} \cdot \ln\left(\frac{p_F^2/2m}{\hbar\omega}\right) \simeq 4.13\cdot 10^{17} \quad (\text{sec}^{-1}) \tag{20}$$

By using (6), the value of $q$ by (20) and $\omega = 10^{14}$ we get for this example

$$\omega_p^2 = \frac{ne^2}{\varepsilon_0 m} = 3.21\cdot 10^{34}\ (\text{sec}^{-2})\ ;\ \varepsilon_1 \simeq 1 - \frac{\omega_p^2}{q^2 Z_i^4} = 1 - \frac{0.19}{Z_i^4}\ ;$$

$$\varepsilon_2 = \frac{\omega_p^2}{\omega q} \simeq \frac{0.78\cdot 10^3}{Z_i^2}\ ;\ \alpha \simeq \frac{\omega_p^2}{cq\eta} = \frac{2.59}{Z_i^2 \eta}\cdot 10^8 \tag{21}$$

We find in this example that the real value of the dielectric constant has only a small difference with 1 but the imaginary part has a large value. The index of refraction $n$, according to (8), has approximately the value $\sqrt{\frac{\varepsilon_2}{2}}$ so that the absorption coefficient $\alpha$ shows a strong decay of incident electromagnetic field.

## 5. The stability of cold stars plasma and the mass and radius of such stars

While stabilities of white dwarfs were studied under relativistic conditions our interest in the present paper is to study this problem in the non-relativistic region, for medium temperatures and high electrons densities. Let us describe the equations for the stability of star in the non-relativistic region.

The general expression for the pressure of the dense plasma under the Fermi-Dirac statistic can be given as

$$P = \frac{1}{3}\frac{2}{h^3}\int pv f(\vec{x},\vec{p}) d^3 p \ . \tag{22}$$

Here $v$ is the electron velocity, the factor 1/3 follows from the hypothesis of isotropy and the factor 2 is due to the two spin states of the electron. This equation defines the pressure as the momentum flux. By using the approximation of (2) we get:

$$P = \frac{8\pi}{3h^3}\int_0^{p_F}\frac{p^4}{m}dp = \frac{8\pi}{15h^3}\frac{p_F^5}{m} \ . \tag{23}$$

If there are $\kappa$ nucleons for each electron the mass density $\rho$ is given approximately by

$$\rho = \kappa n m_N \ . \tag{24}$$

Here $m_N \simeq 1.67\cdot 10^{-27} kg$ is the mass of the nucleon. Then $p_F$ of (3) is given as

$$p_F = h\left(\frac{3}{8\pi\kappa m_N}\rho\right)^{1/3} \quad . \tag{25}$$

By substituting (25) in (23) we get

$$P = \left[\left(\frac{h^2}{5m}\right)\left(\frac{3}{8\pi}\right)^{2/3}\left(\frac{1}{\kappa m_N}\right)^{5/3}\right]\rho^{5/3} = K\rho^{5/3} \quad . \tag{26}$$

The constant K in (26) is defined by the expression in the square brackets. The pressure $P$ depends only the density $\rho$ (within the approximation (2)) and its direct dependence on temperature $T$ vanishes [Chandrasekhar 1967; Eddington 1959].

The stability of a star with spherical symmetry has been studied under the opposing forces of gravitation and pressure in the absence of rotation and other disturbing causes, e.g. magnetic field etc. The gravitational force at a distance r from the center of the sphere is given by

$$g = GM_r / r^2 \quad . \tag{27}$$

Here G is the constant of gravitation and $M_r$ is the mass interior to $r$. g is related to the potential $\phi$ as

$$g = -\frac{d\phi}{dr} \quad . \tag{28}$$

Poisson equation is obtained for spherical symmetry which has the form

$$\frac{d^2\phi}{dr^2} + \frac{2}{r}\frac{d\phi}{dr} = -4\pi G\rho \quad , \tag{29}$$

where [Chandrasekhar 1967; Eddington 1959]:

$$\rho = \left[\frac{(2/5)\phi}{K}\right]^{3/2} \quad ; \quad P = (2/5)\rho\phi \quad . \tag{30}$$

Equation (29) is transformed into the form:

$$\frac{d^2\phi}{dr^2} + \frac{2}{r}\frac{d\phi}{dr} + \alpha^2\phi^{1.5} = 0 \quad ; \quad \alpha^2 = \frac{4\pi G}{(2.5K)^{1.5}} \quad . \tag{31}$$

The procedure for solving the equations for the star stability will consist of solving $\phi$ as function of $r$. Then, $\rho$ and $P$ are found by using (30).

Let $\phi_0$ be the value of $\phi$ at the star center, and let

$$\phi = \phi_0\theta \quad , \quad r = x/\left(\alpha\phi_0^{1/4}\right) \quad , \tag{32}$$

Then, (31) is transformed to

$$\frac{d^2\theta}{dx^2} + \frac{2}{x}\frac{d\theta}{dx} + \theta^{1.5} = 0 \qquad . \qquad (33)$$

The numerical solution of $\theta$ as function of $x$ can be found in [Mohan Al-Bayati 1980] for the Lane-Emden equation with n=1.5 .

Equation (33) is a universal equation describing the potential relative to its value in the star center i.e. $\frac{\phi}{\phi_0} = \theta$ as function of the variable x. But for cases for which the potential in the star center $\phi_0$ is smaller the dependence on the distance $r$ is extended over longer distances relative to the universal $x$ coordinate, by proportionality constant $\phi_0^{-1/4}$. According to (30), the potential $\phi_0$ in the star center is proportional to $\rho_0^{2/3}$ where the density $\rho_0$ is proportional to the number of electrons per unit volume $n_0$ at the star center. We get therefore that the star radius is proportional to $n_0^{-1/6}$ .

Assuming a typical white dwarf with electron density a $n = 10^{34}$ $(met^{-3})$ at its center then we compare it with the above example in which we have a density of $n = 10^{31}$ $(met^{-3})$, so that the number of electrons in the center of such star are smaller by a factor $10^3$ relative to that of the white dwarf. The radius of the star, according to the above analysis corresponding to the above example, increases by factor $10^{3/6} = \sqrt{10}$ relative to the white dwarf and its volume increases by factor: $10^{3/2}$. So in conclusion the gravitational force outside such star will be smaller by a factor $10^{3/2}$ relative to the white dwarf. But many such stars distributed over large volumes can, however, lead to gravitational forces which will be in the order of that of white dwarf.

## 6. Summary and conclusions

In the present work we have taken into account that Boltzmann statistics should be changed to Fermi-Dirac statistics under the condition that the de-Broglie wavelength is large relative to the average distance between the electrons. For cases for which the chemical potential $\mu$ in units of $kT$ is much larger than 1.5, the general Fermi-Dirac distribution of (1) is simplified into the form of (2). The pressure of electrons given by (3) is independent of temperature as it is a purely quantum effect. These properties are common in the treatment of white dwarfs for which the densities are extremely large and for the present plasmas which are for smaller densities and medium temperatures.

The optical properties of cold plasmas with low densities were treated in section 3. We treated the plasma under the condition that the frequency collisions, is very large relative to the electromagnetic optical and infra-red frequencies. The frequency of collisions was related to electron-ion Rutherford scattering. I have taken into account the condition under which the Rutherford-Born approximation is valid. I developed an equation for the averaged value of the frequency of collisions by integrating it over the Fermi-Dirac distribution which takes into account quantum effects. The lower bound in this integration was related to the effect of many collisions with small changes in energy which leads to certain Coulomb logarithmic function.

We obtained three interesting results: a) By using Fermi-Dirac statistics we can get very dense plasmas with temperatures which are much lower than those obtained by Boltzmann statistics so that their luminosity is relatively low. b) We obtained a new quantum mechanical equation for the frequency of collisions by averaging Rutherford scattering effects over the Fermi-Dirac distribution. The condition under which the Rutherford-Born scattering theory is valid was taken into account. The effect of many weak collisions was taken into account by using a certain Coulomb logarithm function. c) Following conventional methods taken from the field of white dwarfs the universal Lane-Emden equation was used for comparing the radius and mass of typical white dwarf with those of the present plasmas given for lower densities and temperatures.